**Modeling between-farm transmission dynamics of porcine epidemic diarrhea virus: characterizing the dominant transmission routes**

Jason A. Galvis [1], Cesar A. Corzo[2], Joaquín M. Prada[3], Gustavo Machado[1]*

[1]Department of Population Health and Pathobiology, College of Veterinary Medicine, North Carolina State University, Raleigh, North Carolina, USA.

[2]Veterinary Population Medicine Department, College of Veterinary Medicine, University of Minnesota, St. Paul, MN, USA.

[3]School of Veterinary Medicine, Faculty of Health and Medical Sciences, University of Surrey, Guildford, UK.

*Corresponding author: gmachad@ncsu.edu

**Abstract**

The role of transportation vehicles, pig movement between farms, proximity to infected premises, and feed deliveries has not been fully considered in the dissemination dynamics of porcine epidemic diarrhea virus (PEDV). This has limited efforts for disease prevention, control and elimination restricting the development of risk-based resource allocation to the most relevant modes of PEDV dissemination. Here, we modeled nine modes of between-farm transmission pathways including farm-to-farm proximity (local transmission), contact network of pig farm movements between sites, four different contact networks of transportation vehicles (vehicles that transport pigs from farm-to-farm, pigs to markets, feed distribution and crew), the volume of animal by-products within feed diets (e.g. animal fat and meat and bone meal) to reproduce PEDV transmission dynamics. The model was calibrated in space and time with weekly PEDV outbreaks. We investigated the model performance to identify outbreak locations and the contribution of each route in the dissemination of PEDV. The model estimated that 42.7% of the infections in sow farms were related to vehicles transporting feed, 34.5% of infected nurseries were associated with vehicles transporting pigs to farms, and for both farm types, pig movements or local transmission were the next most relevant routes. On the other hand, finishers were most often (31.4%) infected via local transmission, followed by the vehicles transporting feed and pigs to farm networks. Feed ingredients did not significantly improve model calibration metrics, sensitivity, and specificity; therefore it was considered to have a negligible contribution in the dissemination of PEDV. The proposed modeling framework provides an evaluation of PEDV transmission dynamics, ranking the most important routes of PEDV dissemination and granting the swine industry valuable information to focus efforts and resources on the most important transmission routes.

**Keywords:** spatiotemporal model, stochastic model, truck, feed formulation.

**Introduction**

The porcine epidemic diarrhea virus (PEDV) U.S. breeding herd incidence has gradually decreased since 2015 (MSHMP, 2021), despite that PEDV continues to spread across multiple pig-producing companies (Machado et al., 2019; MSHMP, 2021; Perez et al., 2019). A previous study compared three different mathematical models that assessed the role of between-farm pig movements, farm-to-farm proximity, and the continued circulation of PEDV within infected sites, named hereafter as "re-break" on PEDV transmission (Galvis et al., 2021b). Although the transmission routes tested helped to explain between farm PEDV dynamics, the study did not fully consider indirect contacts through between-farm transportation vehicles (e.g. vehicles transporting pigs, feed, or farm personnel) contact networks, which has been previously described as major routes of between-farm pathogen transmission (Büttner and Krieter, 2020; Niederwerder, 2021; Porphyre et al., 2020), such as PEDV (Lowe et al., 2014; VanderWaal et al., 2018), African swine fever (ASF) (Gao et al., 2021; Gebhardt et al., 2021), and more recently for porcine reproductive and respiratory syndrome virus (PRRSV) dynamics (Galvis et al., 2021a).

As described recently (Büttner and Krieter, 2020; Galvis et al., 2021a), pig farms are strongly connected through vehicle networks, which highlight the potential of pigs and vehicles in the propagation of swine diseases (Garrido-Mantilla et al., 2021). Previously, PEDV was found in 5.2% of trailers in which pigs had been transported (Lowe et al., 2014), which suggests contaminated vehicles may be a source of the virus. Other epidemiological studies demonstrated a similar association between transportation vehicles and the propagation of documented PEDV outbreaks (VanderWaal et al., 2018). PEDV is an *Alphacoronavirus* that remains viable for an extended period of time on fomites, for example, PEDV was found after 20 days post-contamination in styrofoam, metal, and plastic at low temperatures (4°C) (Kim et al., 2018) which are common materials found in vehicles utilized in the transportation of feed and pigs. On the flip side, vehicle cleaning and disinfection procedures when performed properly and under high temperatures may successfully inactivate PEDV (Niederwerder and Hesse, 2018). Thus, it is clear that there exists a risk associated with PEDV propagation through contaminated vehicles, and quantifying such unknown risk may provide valuable knowledge to the swine industry to design target control strategies including vehicle movement restrictions and cleaning and disinfection procedures.

Similar to the transmission through vehicles, there is evidence that contaminated feed and/or ingredients may play an important role in the propagation of PEDV (Dee et al., 2020; Huss et al., 2017; Schumacher et al., 2018). In 2013, there was some evidence that PEDV may have been introduced in the U.S. through contaminated feed ingredients (Dee et al., 2014; Snelson, 2014), and then introduced into Canada in 2014 through spray-dried porcine plasma (SDPP) used as a feed supplement (Pasick et al., 2014). However, in these studies feed was indicated as a plausible cause, but never determined as the true source of infection (USDA, 2015). Despite that, it was identified that pelleted feed contaminated with a minimum of $5.6 \times 10^1$ TCID50/g of PEDV can infect pigs (Schumacher et al., 2016). Furthermore, PEDV was found in unopened pelleted feed bags (Bowman et al., 2015). Therefore, it is plausible that even pelleted feed becomes contaminated and can become a source of pathogen transmission (Aubry et al., 2017; Schumacher et al., 2017). Indeed, it has been demonstrated that once PEDV is introduced into feed mills, the organism could widespread within the facility (Schumacher et al., 2017) and infectious material could be transferred to subsequently manufactured feed.

Identifying the routes associated with the PEDV propagation is expected to provide the necessary knowledge to formulate more effective preventive, control and elimination strategies, as well as highlight the necessary data to reproduce the disease transmission dynamics (Salines et al., 2017). In this study, we built on a previously developed and flexible mathematical model used to reconstruct the dissemination of infectious viral diseases in swine populations (Galvis et al., 2021c). Here, we extend the mathematical framework *PigSpread* capabilities and model PEDV transmission using nine modes of between-farm propagation: local transmission by the farm-to-farm proximity, between-farm animal and re-break for farms with previous PEDV outbreaks, with the addition of vehicle movements (feed, shipment of live pigs between farms and to slaughterhouses, and farm personnel (crews) and the volume of animal by-products which was restricted to animal fat, meat and bone meals in pig feed ingredients. The model was used to estimate the weekly number of PEDV outbreaks and their spatial distribution, which were compared to available data, and to quantify the contribution of each transmission route.

**Material and Methods**

*Data sources*

In this study, we used weekly PEDV records collected by pig-producing companies and captured by the Morrison Swine Health Monitoring Program (MSHMP) (MSHMP, 2021). Data included outbreaks between January 01, 2015, and June 01, 2020, from 2,294 farms from three non-commercially related pig production companies (coded as A, B, and C) in a U.S. region (not disclosed due to confidentiality). A list of active pig farms was obtained from each company, which included individual national premises identification number, farm type (sow [which included farrow-to-wean, farrow-to-finish and farrow-to-feeder farms], nursery, finisher [which included wean-to-feeder, wean-to-finish, feeder-to-finish], gilt development unit [which could be either part in finisher or sow farms depending upon farm type used by pig production company], isolation and boar stud), pig spaces which stands for the number of head space for each farm, and geographic coordinates. Between-farm pig movement data from January 01, 2020, to December 31, 2020, were collected directly from the pig production companies' database and used to reconstruct directed weekly contact networks. Each movement of a group of pigs included movement date, farm of origin and destination, number of pigs transported, and purpose of movement (e.g. sow, nursery, weaning). Movement data missing either the number of animals transported, farm type, farm of origin, or destination were excluded prior to any analysis (731 unique movements were excluded). In addition, four networks formed by transportation vehicles were recorded from the global positioning system (GPS) vehicle tracker, which included all farms of company A (76% of all farms within the study region) from January 01, 2020 to 31 December, 2020. These movements comprised real-time GPS records of each transporting vehicle, which include geographic coordinates for every five seconds, of any vehicle. Overall, 398 vehicles were monitored which included: (i) 159 trucks used to deliver feed to farms, (ii) 118 trucks utilized in the transportation of live pigs between farms, (iii) 89 trucks used in the transportation of pigs to markets (harvest plant), and (iv) 32 vehicles used in the transportation of crew members, which by the information we collected corresponded to the movement of additional personnel needed for vaccination, pig loading/unloading among other activities which included power washing barns and other facilities. Each vehicle movement event included a unique identification number, speed, date, and time along with coordinates of each vehicle location recorded every five seconds. We defined vehicle visits using the same methodology from a previous study (Galvis et al., 2021a). Briefly, a valid visit was counted by tracing vehicle coordinates (e.g. latitude & longitude) and the vehicle speed. Thus, a constant was counted when the speed was at zero km/h for at least five minutes and the vehicle coordinate was within a radius of 1.5 km from a farm or a cleaning station (truck wash). In addition, we calculated the time in minutes the vehicle remained within each farm's perimeter and the vehicle contact networks between the farms were built considering the elapsed time a vehicle visited two or more different farms. To accommodate PEDV survivability in the environment, we considered two seasons (cold and warm weather) based on previous literature (Kim et al., 2018). The survivability of PEDV through time was considered to be the same as in our previous study of PRRSV (Galvis et al., 2021a). Therefore, two different farms were considered connected by a vehicle if the time between the two visits was less than 72 hours or 24 hours, for the cold and warm seasons, respectively. Furthermore, edges between two consecutively visited farms were disregarded when any vehicle was identified making a stop at a cleaning station. The connections (edges) for all four vehicle networks were weighted by the elapsed time each vehicle visited two different farms, which was later transformed to a probability assuming a decreasing linear relationship of PEDV stability in the environment (Supplementary Material Figure S1). Additionally, we collected feed loadout records from all (three) feed mills used to feed the entire pig population of company A for 2020, with each feed record including feed mill identification with individualized feed formulation (ingredients), amount of feed delivered, destination farm identification, and destination farm delivery data. From the feed loadout records, we collected the amount in pounds (lb) of animal by-products (parts of a slaughtered animal that included animal fat, pig plasma, and meat and bones meals) of each feed formulation received by the farms throughout 2020. Although company B and C data about vehicle movements and feed delivery was not available, we kept the farms from both companies in the transmission model to complement the PRRSV dissemination by the local transmission (Jara et al., 2020).

**Descriptive analysis**

*Between farm animal and transportation vehicles movement*

In a previous study, we analyzed the networks of live pigs transported between farms and four types of transportation vehicles visiting farms described in more detail elsewhere (Galvis et al., 2021a). Here, we estimated whether farms with PEDV outbreaks were more frequently connected with other infected farms through the ingoing contact chain (ICC) and outgoing contact chain (OCC) (Supplementary Material Table S1), compared with farms without PEDV records using a Mann-Whitney test. Briefly, we define contact chains as the subsets of farms that can reach or be reached by a specific farm through direct contact or indirect contacts using a sequential order of edges in a temporal network (Nöremark and Widgren, 2014). In addition, we estimated the association between the time spent within farms' premises by each transportation vehicle and the increase in the number of PEDV outbreaks through a Mann-Whitney test.

*Animal by-products in feed ingredients*

We calculated the total amounts of animal fat, pig plasma, protein blend (protein blend is typically a combination of ingredients such as meat meal, corn germ meal, hominy and dried distillers grains with solubles), and meat and bone meal (MBM) present within 23 distinct feed formulations. In order to further evaluate the association between PEDV outbreaks and the delivery of feed with animal by-products, we performed a logistic regression analysis for each farm type and ingredient in which the response variable was positive or negative for PEDV from January 1st, 2020 to June 6th, 2020, and the predictor was the amount of animal origin feed ingredients divided by the farm's pig capacity, to accommodate possible effect of farm size (number of pigs to be fed)

*Epidemiological model formulation*

The analysis of the spatiotemporal distribution of farm-level PEDV outbreaks was based on our previously developed model *PigSpread* (Galvis et al., 2021b, 2021c), which here was extended to include vehicle transportation networks and the delivery of animal by-products. The model was calibrated to the weekly PEDV outbreaks and considering nine transmission routes including contact network of discrete pig movements (1); the local transmission events between neighboring farms, farm-to-farm proximity (2); re-break by previous exposure to PEDV (3), indirect contact by vehicles coming into farms, including for feed (4); animal delivery to farms (5) and market (6); and vehicles used by personnel (crew) involved in the loading and unloading of pigs (7); the amount of animal fat (8) and meat and bone meal in feed formulation delivered to farms (9); further describe in Table 1 and Figure 1. The model simulates between farm transmission among three farm-level infectious states, Susceptible(S)-Infected(I)-Outbreak(O), SIO model (Figure 1), and we defined Susceptible status as farms free of PEDV, Infected status as farms with PEDV, but yet not detected infectious pigs and Outbreaks status as infected farm that detected PEDV. Thus, farms in a susceptible state (*i*) receive the force of infection of infected and outbreak farms (*j*) in each time step *t* and become infected at rate $Y_{it}$. The latent period of PEDV is not explicitly modeled, as it is typically a few days after infection, and often viral shedding starts within seven days post-infection (Niederwerder and Hesse, 2018), thus it is embedded in the weekly timestep. Local transmission was modeled through a gravity model, where the probability of infection is proportional to the animal capacity of the farms and inversely related to the distance between two farms (i.e. lower transmission at longer distances), with the maximum distance set at 35 km, similar to our previous study to facilitate the comparison of results (Galvis et al., 2021b). Local transmission is also dependent on the enhanced vegetation index (EVI) around each farm $i$ (Jara et al., 2020), such that the probability of transmission decreases with high EVI values (Supplementary Material Figure S2). The transmission associated with between-farm pig movements is modeled by the number of all infected and outbreak farms sending pigs to susceptible farms. The dissemination via transportation vehicle networks (e.g. vehicles transporting pigs to farms) is modeled by the edge weight ($E$) and the time the vehicles remained on the susceptible farm premise ($Z_{it}$) (Supplementary Material Figure S1 and S3). The transmission via animal fat and meat and bone meal was only considered on sow farms and modulated by the amounts delivered to susceptible farms ($A_{it}$). Pig plasma and protein blend meals were only delivered to nursery and finisher farms, thus were not considered. For the re-break rate, which is only considered for sow farms, we assumed that subsequent new infections at an individual farm, within a time period of two years, were associated with the same strain as the previous outbreak, and the probability was based on a survival analysis evaluating the time farms re-break after recovering ($W_{it}$) (Supplementary Material Figure S4). Then, for each transmission route, the force of infection ($\lambda$) of infected and outbreak farms varies with a seasonality derived from analysis of the PEDV records from 2015 to 2019 (Supplementary Material Figure S5). In addition, sow farms without a record of PEDV outbreaks since 2015 were assumed to have high biosecurity levels ($H$) that reduce the force of infection received by infected and outbreak farms, being $H$ higher than zero and calibrated according to the observed outbreaks (Supplementary Material Table S2). Otherwise, farms with outbreaks records were assumed to have low biosecurity levels and $H$ was defined as zero.

The transition from infected to an outbreak status is estimated through a detection rate $f(x)$. Thus, the probability that farms going from susceptible into outbreak state was dependent on the maximum detection probability ($L$), considered equal to cases reported to MSHMP (MSHMP, 2021), and the average time it takes a farm to detect the disease ($x0$), assumed to be three weeks, estimated from information provided by local swine veterinarians and previous literature (Niederwerder et al., 2016). The proportion of Infected and Outbreak sow farms that return to a susceptible state is drawn from a Poisson distribution with a mean of 28 weeks, which is the average time to stability (Goede and Morrison, 2016). Nursery and finisher farms' transition to susceptible status is driven by pig production movement scheduling of the all-in all-out management schemes of closeouts or by incoming or outgoing movements, whichever came first (Galvis et al., 2021c). Briefly, nurseries and finisher farms become susceptible within 7 and 25 weeks of pig placement, respectively, or when at least one new pig movement is recorded before the farm reaches the scheduled production phase timeline described earlier. A detailed description of the model can be found in previous work describing in greater detail other model parameters (Galvis et al., 2021c). Finally, we used an Approximate Bayesian Computation (ABC) rejection algorithm (Beaumont, 2010; Minter and Retkute, 2019), to estimate the posterior distribution of unknown model parameters (list of model parameters available in Supplementary Material Table S3) by selecting the particles that best fitted the temporal and spatial distribution of observed PEDV outbreaks (Supplementary Material Section 2).

*Model outputs*

The model outputs were derived from a random sample of the 100 accepted particles in the ABC rejection algorithm (number particles accepted defined according to our computational resources). The model outputs included (a) the force of infection for each farm type and transmission route, (b) the weekly number of infected undetected and detected farms (outbreaks), and (c) the sensitivity to detect PEDV outbreaks locations (Supplementary Material Section 2). We carried out 1,000 runs to estimate the relative contribution of each transmission route and weekly number of cases, while for the model sensitivity performance we only used 100 interactions due computational resources (additional information see Supplementary Material Figure S8). Here, we defined the contribution of the routes as the weekly number of infected farms resulting from each transmission route individually, which were then divided by the weekly number of simulated infected farms from all the combined routes and commercial companies. In addition, the average contribution for each route was estimated by summing the weekly contributions divided by the number of simulated weeks, and credible intervals (CI) using an equal-tailed interval (ETI) method were estimated from the weekly contribution distribution. The model was developed in the R (3.6.0 R Core Team, Vienna, Austria) environment, and all simulations were run in RStudio Pro (1.2.5033, RStudio Team, Boston, MA) and transmission model framework is available at https://github.com/machado-lab/pigspread.

**Results**

We evaluated the association of PEDV outbreaks frequency within both ICC and OCC from infected and non-infected farms. The results showed that PEDV infected farms were more frequently found within the ICC and OCC of other infected farms in five out of the six networks ($p < 0.05$), the exception was the vehicles transporting feed in which we only found significant association for ICC ($p < 0.05$) (Supplementary Material Figures S9 and S10). We also evaluated the association between the time transportation vehicles remained within infected and non-infected farms. The vehicles transporting pigs to farms spent more time within infected nursery farms when compared with the non-infected farms ($p < 0.05$), vehicles transporting pigs to market spent more time within infected nursery farms when compared with the non-infected farms ($p < 0.05$), and vehicles transporting crew spent more time within the infected nursery and finisher farms when compared with the non-infected farms ($p < 0.05$) (Supplementary Material Figures S11). In addition, no differences were found among the time spent on PEDV infected farms compared with the non-infected farms for the vehicles transporting feed for any farm type ($p > 0.05$). Finally, for all animal by-products in feed ingredients, our results indicated that nursery farms with previous outbreaks reports received higher amounts of pig plasma and meat and bone meals compared with farms without outbreaks reports, and this difference was significant using logistic regression ($p < 0.05$) (Supplementary material Figures S12-S14). On the other hand, other ingredients (animal fat and protein blend) were not significantly associated with PEDV outbreaks for any farm type ($p > 0.05$) (Supplementary Material Figures S12-S14).

From the epidemiological model simulations, we estimated the average number of infected farms which was in total 651 (95% CI: 646–657), 39 (95% CI: 38–40) of which corresponded to infected sow farms, 200 (95% CI: 197–203) to nursery farms and 412 (95% CI: 409–415) to finisher farms (isolation and board stud farm were excluded from results as no outbreaks were reported in the period studied). Overall, results showed a satisfactory agreement between the weekly observed number of PEDV outbreaks and simulated outbreaks (Supplementary Material Figure S7). The model inferred that at the end of the 21 weeks, on average 74 (11.3%) of all PEDV infected farms would be detected in which 90% of the infected sow farms were detected, while a much lower proportion of detection was estimated for nurseries (6.5%) and finishers farms (6.3%). The predictive performance of the model to correctly identify the weekly spatial distribution of known PEDV outbreaks showed an area under the curve (ROC) of 0.86 (Supplementary Material Section 2).

The contribution of nine transmission routes in the PEDV spread over time is available in Figure 2. Evaluating the average contribution, for company A's farms our results revealed that for sow farms the most important route was the vehicles transporting feed as it contributed to an average of 42.7% (95% CI 7%-73%) of the farm infections, followed by local transmission with 19.4% (95% CI 5%-41%), vehicles transporting pig to farms 14.9% (95% CI 0.8%-41%), pig movements 13.1% (95% CI 0.2%-57%), re-break 5.4% (95% CI 1%-12%), vehicles transporting crew 2.8% (95% CI 0%-12%), vehicles transporting pigs to markets 1.7% (95% CI 0%-13%), and both amounts of animal fat and amount of meat and bone meals within feed formulation were at 0% since the inclusion of such route did not contribute to sensitivity during model calibration (Figure 2). For nursery farms, vehicles transporting pigs to farms were the most important route contributing to 34.5% (95% CI 7%-55%) of the farm infections, followed by pig movements 26.1% (95% CI 0%-50%), local transmission with 22.6% (95% CI 10%-54%), vehicles transporting feed 13.7% (95% CI 7%-25%), vehicles transporting crew 1.8% (95% CI 0%-7%), and vehicles transporting pigs to markets 1.3% (95% CI 0%-6%). For finisher farms, local transmission was also the most important route contributing to 31.4% (95% CI 16%-58%) of the farm infections, followed by vehicles transporting feed with 29.6% (95% CI 13%-44%), vehicles transporting pigs to farms 21.8% (95% CI 5%-35%), pig movements 8% (95% CI 0%-30%), vehicles transporting pigs to markets 6% (95% CI 0.7%-12%) and vehicles transporting crew 3.1% (95% CI 0.6%-7%). Of note, transportation vehicle data were not available for companies B and C, the results were restricted to three transmission pathways (pig movements, local transmission, and re-break) and are available in Supplementary Material section 4, Figure S15.

**Discussion**

Our results quantified the contribution of nine transmission pathways in the dissemination dynamics of PEDV which included pig movement network, farm-to-farm proximity, re-break, different types of transportation vehicle networks (vehicles transporting feed, pigs to farms, pigs to markets, and crew), and the delivery of animal by-products, in particular, animal fat and meat and bone meal in the feed. Our results demonstrate that vehicles transporting feed to farms were the most important route infecting sow farms, while vehicles transporting pigs to farms were the most important for nursery farms, and for both farm types, pig movements or local transmission were the next most relevant routes of PEDV dissemination. On the other hand, finishers were more often infected via local transmission, followed by vehicles transporting feed and pigs to farm networks (Figure 2). The volume of animal fat and meat and bone meals in the dynamics of PEDV did not significantly improve model calibration metrics, sensitivity, and specificity; therefore, between-farm dissemination seems to be independent from the evaluated by-products in the studied period of time.

From our descriptive analysis, infected farms were more frequently located within the ingoing and outgoing contact chain of other infected farms. In most comparisons, there was a negligible difference in the number of infected farms in the contact chain among the farms with and without reported PEDV outbreaks (Supplementary Material Figures S9 and S10), and probably the test power was influenced by the large number of farms analyzed (n = 2,294), which makes it more likely to find difference among the compared groups. On the other hand, as described elsewhere (Trevisan et al., 2021), surveillance and disease reporting at downstream farms (nurseries and finishers) is not systematic as at breeding farms (sows); therefore, it is possible that the significant differences for the number of cases in each contact chain are likely to be much greater than what was observed. The transmission of disease through contact chains from pig movement networks is especially relevant in vertically integrated systems such as in areas of high commercial pig production like in North America (Galvis et al., 2021c; Passafaro et al., 2020; Rautureau et al., 2012), in which the continuous flow of pigs moving from breeding sites into finish sites are heavily utilized. On the other hand, there is limited information about the contact chain pattern of transportation vehicle movements and the dissemination of swine diseases between-farms (Büttner and Krieter, 2020). Therefore, in-depth analysis of vehicle networks is needed in order to uncover more details about their role in the dissemination of swine diseases.

In this study, vehicles transporting feed were closely related to positive sow farms, although there is limited information regarding the mechanistic role of vehicles as fomites for PEDV transmission, a recent outbreak investigation in Mexico suggests that either a contaminated vehicle transporting feed or feed was responsible in the introduction of PEDV into a sow farm (Garrido-Mantilla et al., 2021). Furthermore, another study in the U.S. detected PEDV RNA on the surfaces of a feed truck (Elijah et al., 2022). Through exploration of our model calibration approach, we identified that PEDV outbreaks reported for weeks 6 and 19 showed higher ROC values when the model was calibrated with vehicles transporting feed (Supplementary Material Figure S8). In addition, it is important to note that the network formed by vehicles transporting feed was highly connected (Galvis et al., 2021a), and along with local transmission these two routes may also be filling the gaps of transmission routes not included in this study, such as rendering trucks, people and wildlife (Jung et al., 2020; Niederwerder and Hesse, 2018). Similar to a study evaluating PRRSV transmission dynamics (Galvis et al., 2021a), transmission via vehicles transporting pigs to farms was frequent to all farm types, in which in percentage this route contributed with 14.9% of cases in sow farms, 34.5% nurseries and 21.8% finishers. This highlights the spreading potential of vehicles transporting pigs, which has been described elsewhere (Büttner and Krieter, 2020; Niederwerder and Hesse, 2018; Thakur et al., 2016). The relationship between PEDV outbreaks and contact networks of vehicles transporting pigs to market and loading crews were the least relevant route of transmission, with finisher sites showing the highest proportion of infections by either vehicle transporting pigs to market (6%) or loading crews (3.1%). Similar results have been described in a recent PRRSV epidemiological modeling work (Galvis et al., 2021a), in which both contact networks had limited contribution to the disease propagation.

We demonstrated by descriptive analysis that PEDV positive nursery farms received slightly higher amounts of plasma and meat and bone meal when compared with negative ones. However, similar to the significant associations found in the sizes of contact chains, the large number of farms analyzed may have contributed to improving the power of the statistical test. In addition, the large amount of animal by-products associated with infected farms could be linked with nutritional and management decisions, in which production managers may identify below average weight gains and increase/improve feed formulations to improve feed intakes, thus this result may not be directly associated with disease or the herd health status. The amount of animal fat and meat and bone meal did not significantly affect the model calibration performance; therefore, the hypothesis is that in this study, animal by-products did not contribute to PEDV dissemination dynamics. Nevertheless, we cannot ignore the risk associated with contamination of complete feed as a likely transmission route for PEDV and other swine diseases (Dee et al., 2020). In fact, the recent outbreak investigation in Mexico mentioned above concluded that contaminated feed or a contaminated vehicle transporting feed was the most probable route of infection after identifying the presence of PEDV RNA in the lactating feed, the interior walls of the feed bins, and in samples collected from the interior of the auger boom of the feed truck (Garrido-Mantilla et al., 2021). Although adequate pelleting procedures involve high temperatures and pressure, which have been effective in inactivating PEDV (Cochrane et al., 2017), the manufactured feed can still become contaminated by direct contact with contaminated surfaces within the feed mill facility, thus still carrying viral particles into swine productions (Schumacher et al., 2017). Thus, in order to better understand the role of feed in the dissemination of PEDV and other diseases, more information regarding contamination of feed mill facilities, trucks utilized in feed transportation, and at farm level and in-depth evaluation of contaminated feed bins is needed (Dee et al., 2020; Niederwerder and Hesse, 2018).

**Final remarks and limitations**

We remark on some aspects of this study that may affect our modeling results. First, the model was calibrated to the observed cases in space and time from the available data (January 01, 2020 until June 01, 2020). Thus, the results may change with the inclusion of additional PEDV outbreak data, in addition to other transmission routes (e.g. rendering network), control strategies (e.g. immunization programs) or the inclusion of vehicle routes from companies B and C that were not available. Another limitation to consider was that we defined a vehicle visiting a farm when this was located within a radius distance of 1.5 km from a farm, which in some cases could include more than one farm, thus increasing the number of contacts created by the vehicles. Although we assumed farms within 1.5 km radius from any vehicle were at risk of transmission, a future alternative would be to collect the geolocation of each farms' feed bins and measure the distance to trucks that have stopped within a much shorter distance to the barns. In relation to the model parameters, we simplified some of them, such as on-farm biosecurity, that was considered either high or low according to the historical records of PEDV outbreaks. Thus, future approaches may benefit from the inclusion of more detailed information on biosecurity infrastructure (e.g. cleaning and disinfection stations) and practices (e.g. the type of disposal of dead animals) (Sykes et al., 2021). In this study, we assumed that not all the PEDV outbreaks in downstream farms (nursery and finisher) were reported given the current surveillance programs, which prioritize routine sampling at breeding sites only. Therefore, we assume that the yearly farm level prevalence of PEDV infection at downstream farms was assumed to be only 25% of the total expected number of infected farms (personal communication by consulting with multiple swine veterinarians of companies A, B and C). This approach was similar to what we have proposed in our recent work on PRRSV (Galvis et al., 2021a, 2021c), where we assumed a 30% farm prevalence in downstream farms. Indeed, the current modeling approach differs from our previous study modeling PEDV (Galvis et al., 2021b), in that the model calibration assumed a prevalence considering only the total number of PEDV outbreaks reported in the studied period. Here, the inclusion of the PEDV infection prevalence in downstream farms clearly improved model calibration, which has been evaluated by comparing each model performance at reproducing the temporal trajectories of PEDV only with the observed cases. Hence, we tested several prevalence values and 25% better approximated the model simulation with the observed weekly PEDV outbreaks. Despite the improvements made to the model, given the lack of studies estimating PEDV prevalence, it is expected that model results would need to be updated accordingly when new data becomes available. Finally, we remark that this study is the first to have used nine transmission routes with observed/real data, which offers an opportunity to evaluate the effect of transmission routes rarely considered, such as transportation vehicles and feed delivery.

**Conclusion**

We significantly expanded previous PEDV dissemination modeling attempts by accounting for nine different routes of between-farm transmission, including the role of animal by-products delivered via feed and four transportation vehicle networks. Collectively, these results reinforce the hypothesis that PEDV dissemination is also driven by vehicle movements which interconnect large numbers of farms. Among the vehicle networks analyzed, by far vehicles transporting feed and pigs to farms were the dominant routes. The volume of animal by-products delivered to farms via feed did not contribute to explaining the spatial distribution of PEDV outbreaks. Nevertheless, we highlighted the need for better data on feed contamination within the entire feed manufacturing and transportation chain starting from feed ingredient delivery to the feed mill all the way to delivery into the farm feed bins, before more conclusions can be drawn regarding the role of feed in the transmission of swine disease. Our findings are especially useful for future studies to be performed in other regions or for other swine disease pathogens, wherein decisions about data collection or which mode of transmission should become a priority during disease control and elimination.

**Acknowledgments**

The authors would like to acknowledge participating companies, veterinarians and the Morrison Swine Health Monitoring Project. The Morrison Swine Health Monitoring Project is funded by the Swine Health Information Center.

**Authors' contributions**
JAG and GM conceived the study. JAG, JMP, and GM participated in the design of the study. CC coordinated the disease data collection through the Morrison Swine Health Monitoring Project (MSHMP). JAG and GM conducted data processing, cleaning, designing the model, and simulated scenarios. JAG and

GM designed the computational analysis. JAG and GM wrote and edited the manuscript. All authors discussed the results and critically reviewed the manuscript. GM secured the funding.

**Conflict of interest**

All authors confirm that there are no conflicts of interest to declare

**Ethical statement**

The authors confirm the ethical policies of the journal, as noted on the journal's author guidelines page. Since this work did not involve animal sampling nor questionnaire data collection by the researchers, no IACUC protocol was generated.

**Data Availability Statement**

The data that support the findings of this study are not publicly available and are protected by confidential agreements, therefore, are not available.

**Funding**

This project was funded by the Fats and Proteins Research Foundation.

3 entries

**List of tables**

**Table 1.** List and description of the 9 transmission routes used to model PEDV transmission model.

| Transmission routes | Description |
|---|---|
| **Pig movements** | Transmission route associated with the movement of live pigs between farms. |
| **Local transmission** | Transmission route associated with the distance between farms, as the distance between infected and susceptible farms decrease, the indirect transmission likelihood among them increases. |
| **Vehicles transporting feed*** | Transmission route associated with contaminated surfaces of vehicles transporting feed to farms. |
| **Vehicles transporting pigs to farm*** | Transmission route associated with contaminated surfaces of vehicles transporting live pigs between farms (e.g. wean pig from sow farms into nurseries). |
| **Vehicles transporting pigs to market*** | Transmission route associated with contaminated surfaces of vehicles transporting live pigs from farms to slaughterhouses (market). |
| **Vehicles transporting crew*** | Transmission route associated with contaminated surfaces of vehicles that transport caretakers, service and maintenance personnel which included but not limited to activities such as loading and unloading pigs. |
| **Feed delivery, meat and bone meal** | Transmission route associated with the volume of contaminated meat and bone meal ingredients in feed delivery to farms. |
| **Feed delivery, animal fat** | Transmission route associated with the volume of contaminated animal fat ingredient in the feed delivery to farms. |
| **Re-break** | Probability of the farm re-break by a past exposure to PEDV according to historic outbreaks records based on a survival analysis (Supplementary material Figure S4). |

*For all transportation vehicles routes, we considered that the vehicles were contaminated after visiting an infected farm from which PEDV was carried into susceptible farms.

**List of figures**

**Figure 1. Model framework and transmission routes.** The model simulate the disease spread among three farm-level infectious states, (S) Susceptible-(I) Infected-(O) Outbreak, SIO model using 9 transmission routes: 1) Pig movements network, 2) Local transmission, 3) Re-break, 4-7) four vehicle movement networks and 8-9) volume of two animal by-product ingredients in feed delivered to the farms.

**Figure 2. Farm infection contribution for each transmission route of each farm type (rows).** The *y*-axis represents the proportion of transmission by each transmission route, while the x-axis shows each week in the simulation. Weekly proportions of transmission were calculated by dividing the number of simulated infected farms for each transmission route by the number of simulated infected farms by the total number of routes combined. Of note, animal fat and meat and bone meal contributions were at zero percent.

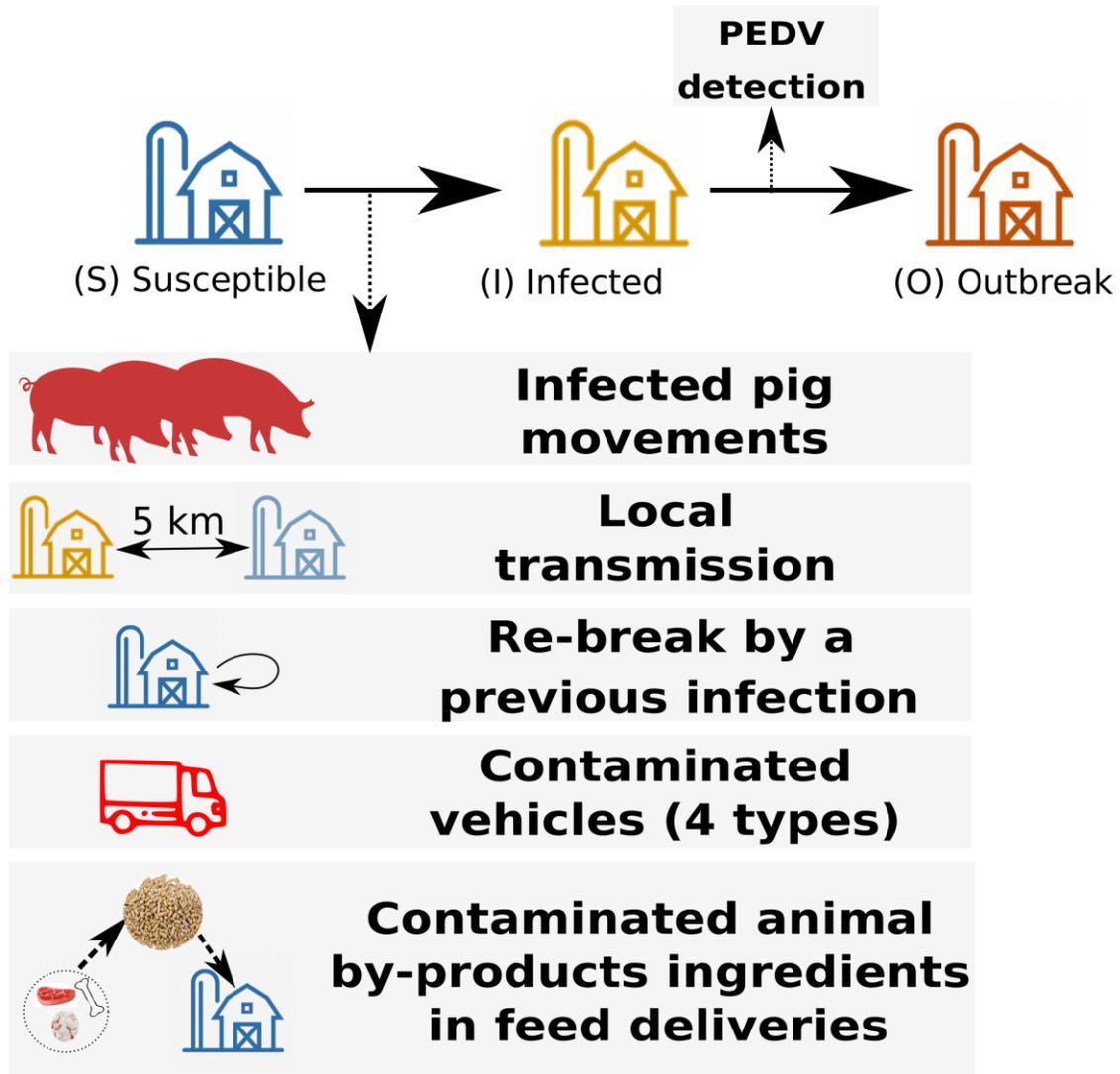

**Figure 1.**

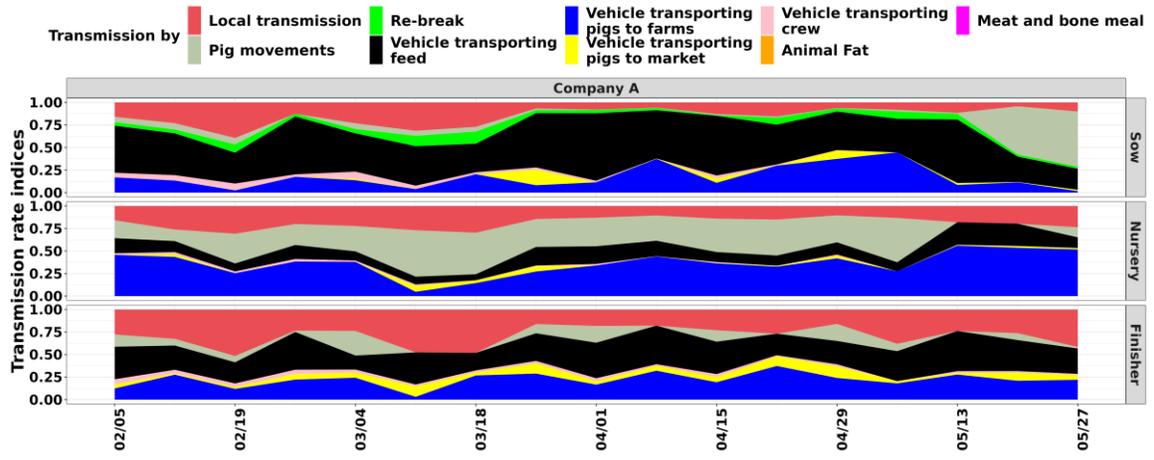

Figure 2.

# Supplementary Material

**Section 1.** A descriptive analysis of model parameters used in the PEDV simulated transmission.

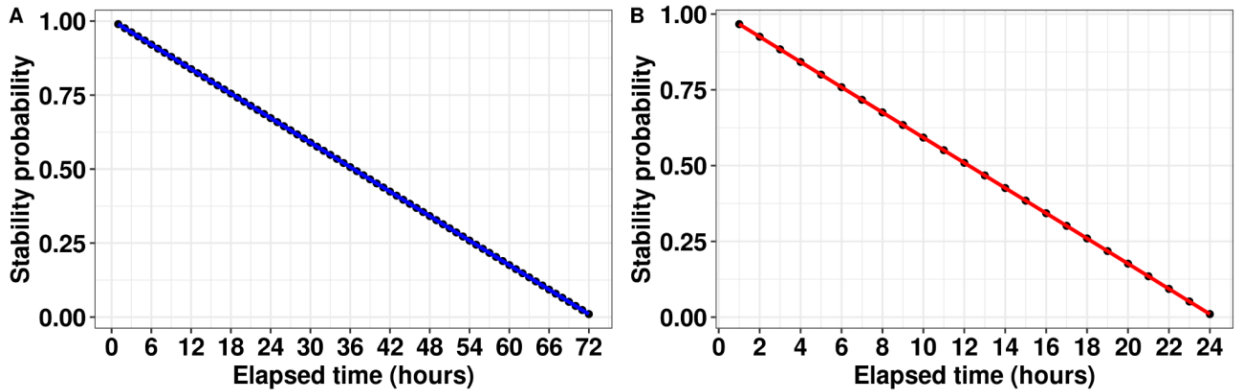

**Figure S1**. The distribution of PEDV stability probability on vehicle surface over time. In A), we show the decay of cold season which included the months between October to March, and B) warm season decay in PEDV suitability which included the months between April to September. In summary, here we assumed that PEDV suitability decreased linearly over time, in which for cold months after 72 hours the suitability of PEDV was set to zero, the same applied to warm weather in which at hour 24 after after a truck visited an infected farm PEDV was no longer viable. This approach was used for all vehicle related contact networks including vehicles transporting feed, pigs-to-farm, pigs-to-market and crew.

**Table S1.** Terminology and definition from the social network analysis.

| Network terminology | Definition | Reference |
|---|---|---|
| Node | Element of the network representing the farms. | - |
| Edge | Link among two nodes. | - |
| Static network | Once an edge exists between two nodes, it is present for the whole time period. | (Kao et al., 2007) |
| Temporal network | The edges between two nodes only exist at different time steps. | (Lentz et al., 2016) |
| Ingoing contact chain (ICC) | Subsets of nodes that can reach a specific node by direct contact or indirect contacts through a sequential order of edges through other nodes using the temporal network. | (Nöremark and Widgren, 2014) |
| Outgoing contact chain | Subsets of nodes that can be reached by a specific node by direct contact or indirect contacts through a sequential order | (Nöremark and Widgren, 2014) |

| (OCC) | of edges through other nodes using the temporal network. | |

To calculate the barrier index (vegetation level, utilized to modulate the probability of local transmission), we used a linear regression, to express PEDV infected farms from 2020 as a function of the Enhanced Vegetation Index (EVI) and yearly seasonality (spring, summer, fall, and winter). We found that PEDV frequency decreased as EVI increased, with a stronger association in winter and fall seasons (Figure S2). Here we used the regression coefficients to predict weekly PEDV incidence, which then was transformed into parameter *a*, which was scaled into values between [0, 1], later utilized to modulate the local transmission.

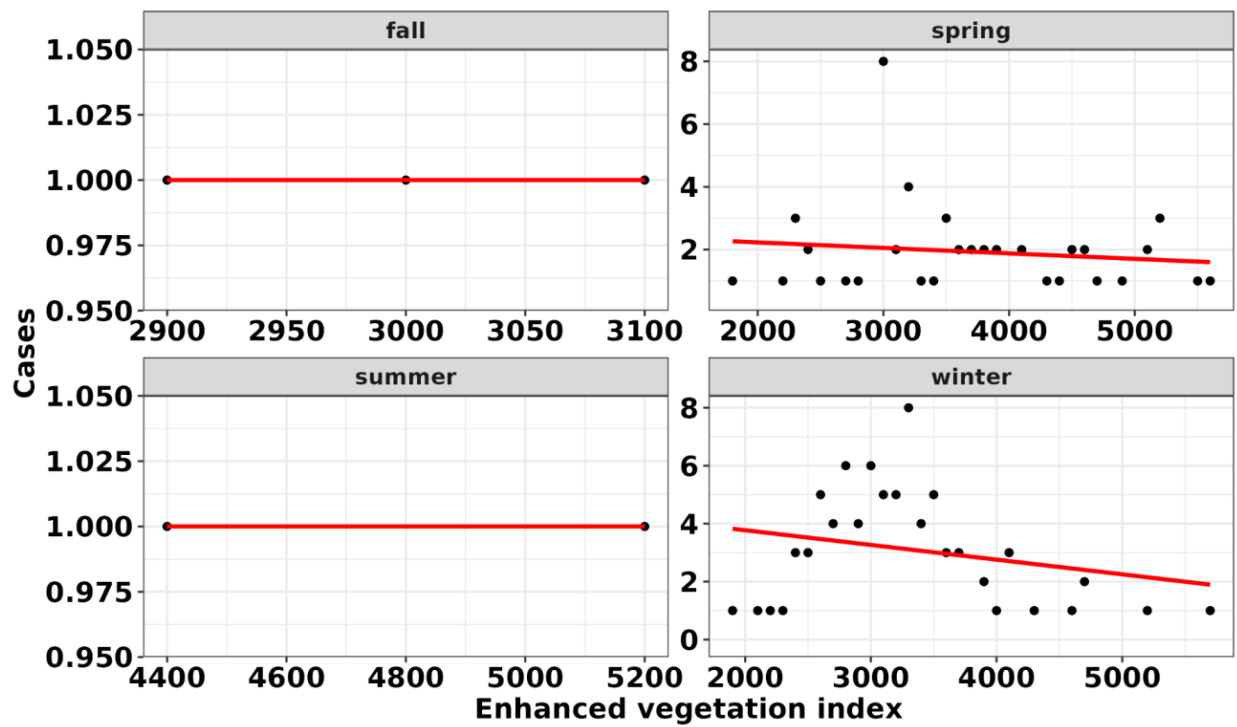

**Figure S2**. Linear regression of PEDV infected farms. The y-axis is the number of PEDV outbreaks and in the x-axis EVI.

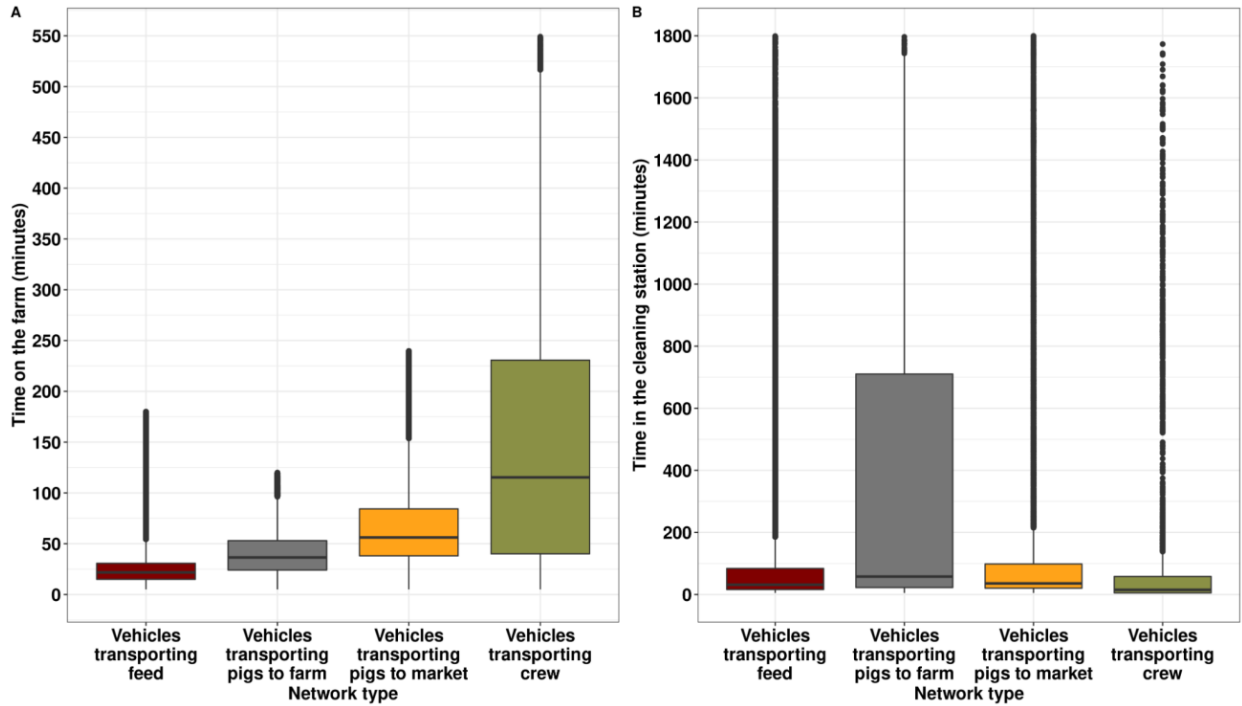

**Figure S3**. The time vehicles spent on each farm visit. The boxplot shows the distribution in minutes that each vehicle remained within farms premises in A) and at cleaning stations in B).

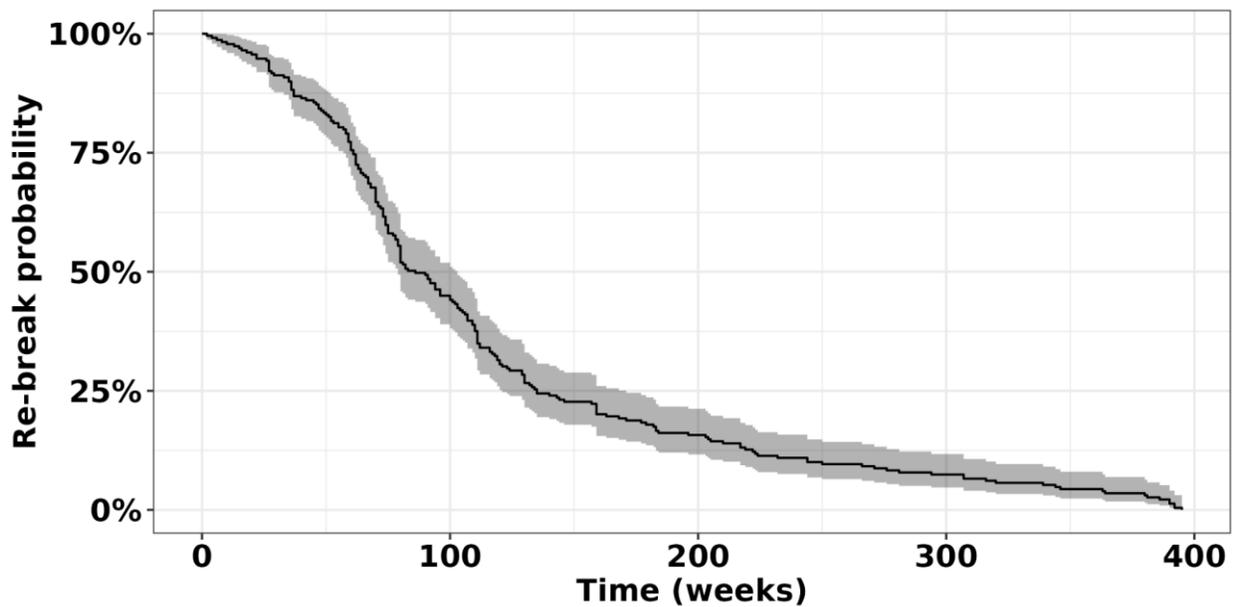

**Figure S4**. A survival analysis of infected and recovered sow farms from PEDV between 2015 and 2019. In this example, we showed the distribution used for each farm to calibrate the re-break probability.

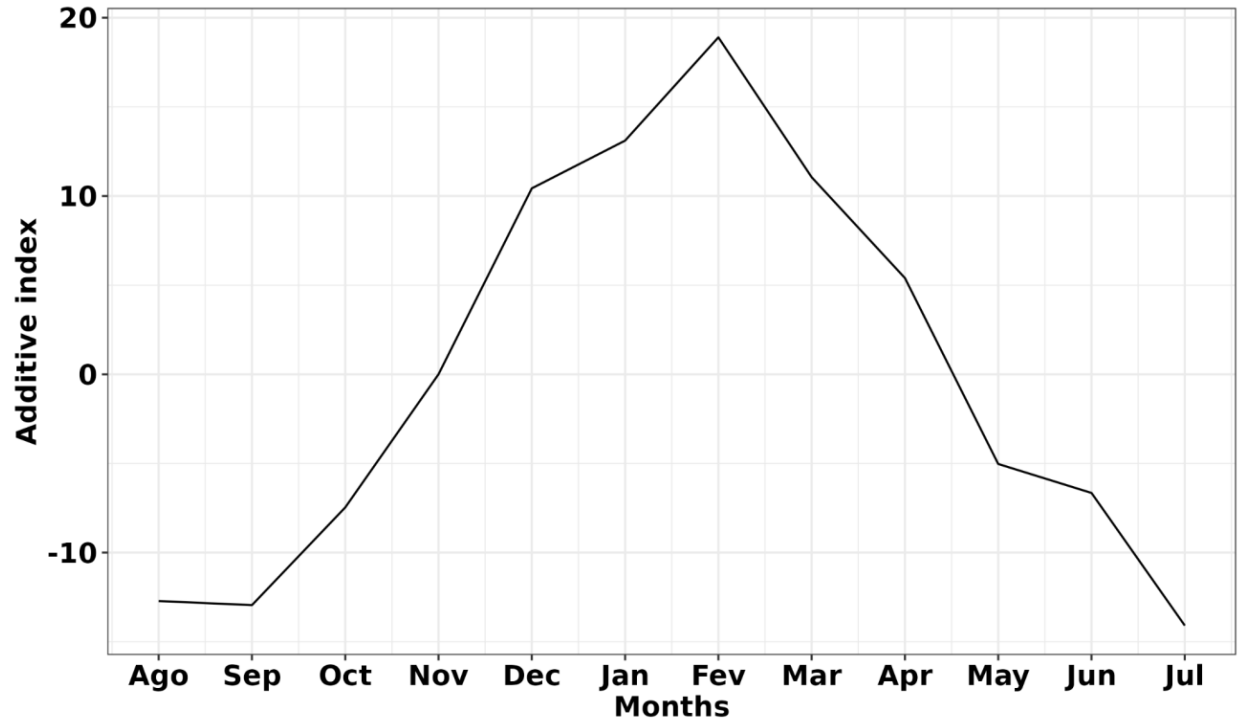

**Figure S5**. Monthly seasonality index calculated from the frequency of PEDV, calculated by an additive moving average decomposition derived from analysis of the PEDV records from 2015 to 2019.

**Section 2**: Model calibration and main model outputs

The model is calibrated in two steps, first fitting the frequency of observed outbreaks on time, and second fitting the location of the observed outbreaks. In Table S2 we show the summary statistics used in step 1 of the Approximate Bayesian Computation (ABC) rejection algorithm, where the tolerance interval represents the square error allowed from the simulation to the observed values. In our model scenario, we considered that PEDV outbreaks were under-reported according to a discussion with local veterinarians about the notification of PEDV infections. Thus, we considered a prevalence of 25% infected farms per year, which was the prevalence that fitted better the ABC Sequential Monte Carlo rejection algorithm after some run tests. In the model calibration, a particle is defined as a set of transmission parameters (Table S3). For the model calibration in step 1, we defined as an accepted particle the set of transmission parameters, whose simulation results are lower than the tolerance interval for all the variables when compared with the observed values. Thus, the range of error accepted (tolerance interval) defines the number of particles accepted in the step 1 of the model calibration. In addition, low tolerance interval values produce particles with similar results to the observed values, but at the same time, the number of particles accepted decreases. Thus, in this study the tolerance interval values were chosen through initial exploratory analysis, selecting the minimum tolerance interval values to accept an adequate number of particles to be analyzed in the step 2 of the model calibration.

**Table S2.** Summary statistics used by the ABC Sequential Monte Carlo rejection algorithm for the model.

| Summary statistics | Observed values | Tolerance interval ($\epsilon$) |
|---|---|---|
| Total number of sow farms with detected cases | 15 | 30 |
| The weekly average number of sow farms with detected cases | 0.7 | 1 |
| The weekly maximum number of sow farms with detected cases | 3 | 5 |
| Total number of nursery farms with detected cases | 31 | 40 |
| The weekly average number of nursery farms with detected cases | 1,4 | 2 |
| The weekly maximum number of nursery farms with detected cases | 9 | 20 |
| Total number of finisher farms with detected cases | 74 | 40 |
| The weekly average number of finisher farms with detected cases | 3.5 | 1 |
| The weekly maximum number of finisher farms with detected cases | 25 | 10 |
| Expected prevalence in finisher and nursery farms | 25% (model assumption) | 100 |

To assess the model performance, we evaluated the probability to predict cells (10 x 10 km squares) with truly infected cells (cells where at least one sow farm outbreak was recorded) at time $t$. Each sow farm was allocated to a cell in the spatial grid (a total of 154 cells in the study area). The risk of each cell was calculated by the sum of times at least one farm within a cell was identified with infected status after 100 simulations based on the distribution of the estimated risk values; we utilized a percentiles thresholds (r) approach to determine cells at high and low risk. Where high-risk cells were compared with the truly infected cells at time $t$. Subsequently, we estimate the model sensitivity and specificity, for all thresholds, as follows:

$$S_r = TP_r/(TP_r + FN_r)$$
$$E_r = TN_r/(TN_r+FP_r)$$

where true positives (TP) was the subset of cells with observed outbreaks and the estimated risk was above the r threshold; false negatives (FN) was the subset of cells with observed outbreaks and the estimated risk was below the r threshold; true negative (TN) was the subset of cells without observed outbreaks and the estimated risk was below the r threshold, and false positives (FP) was the subset of cells without observed outbreaks and the estimated risk above the r threshold. It is worth noting that cells with zero risk were not considered in the sensitivity analysis.

In step 2 of the ABC rejection algorithm, the sensitivity and specificity were calculated for each particle accepted in step 1 of model fitting. The particles accepted were those with sensitivity values ≥50% with r = 85th and ≥70% with r = 70th (percentile and sensitivity thresholds values were chosen arbitrarily after authors discussed the minimum performance of the model). The prior for each parameter were drawn from a uniform distribution that ranged between 0 and 1.5 for pig movements transmission rate, 0 and 0.001 for local transmission, the four transporting vehicles and amount of animal fat and meat and bone meal in the feed meals transmission rates, and finally between 0 and 0.01 for re-break transmission rate. These range values were chosen according to model performance to fit the temporal and spatial distribution of PEDV cases through some test simulations, thus reducing the number of simulations and processing time in the model calibration.

**Table S3.** Transmission parameters are used in simulations, for the distribution of the posterior parameter.

| Model parameter | Symbol | Average values | 95% credible interval* | Details & references |
|---|---|---|---|---|
| The transmission rate of pig movements | $\beta_n$ | 0.72 | 0.1-1.4 | ABC fitting |
| Local transmission rate | $\beta_l$ | 0.00053 | 0.00001-0.001 | ABC fitting |
| The transmission rate of vehicles transporting feed | $\beta_f$ | 0.000013 | 0.000001-0.00003 | ABC fitting |
| The transmission rate of vehicles transporting pigs to farms | $\beta_p$ | 0.00037 | 0.000005-0.0009 | ABC fitting |
| The transmission rate of vehicles transporting pigs to market | $\beta_m$ | 0.00053 | 0.000008-0.001 | ABC fitting |
| The transmission rate of vehicles transporting crew | $\beta_c$ | 0.00056 | 0.0003-0.0009 | ABC fitting |
| The transmission rate of animal fat in the feed meal | $\beta_a$ | 0 | 0-0 | ABC fitting |
| The transmission rate of meat and bone meal in the feed meal | $\beta_b$ | 0 | 0-0 | ABC fitting |
| The transmission rate of re-break | $\beta_r$ | 0.005 | 0.0003-0.009 | ABC fitting |
| Farm' biosecurity | H(sow) | 0.56 | 0.04-0.98 | ABC fitting |
| Maximum effective surveillance | L(sow) | 0.95 | - | Expert opinion |
| | L(nurseries) | 0.070 | 0.04-0.09 | ABC fitting |
| | L(finisher) | 0.0084 | 0.007-0.009 | ABC fitting |
| | L(others) | 0.047 | 0.005-0.09 | ABC fitting |
| PEDV seasonality | T | Weekly values calculated | - | Figure S7 |
| Average time for PEDV detection | X0 | 3 weeks | - | Expert opinion |
| Average infectious time sow farms | - | 28 weeks | - | (Goede and Morrison, 2016) |

*Credible intervals calculated with method equal-tailed interval (ETI).

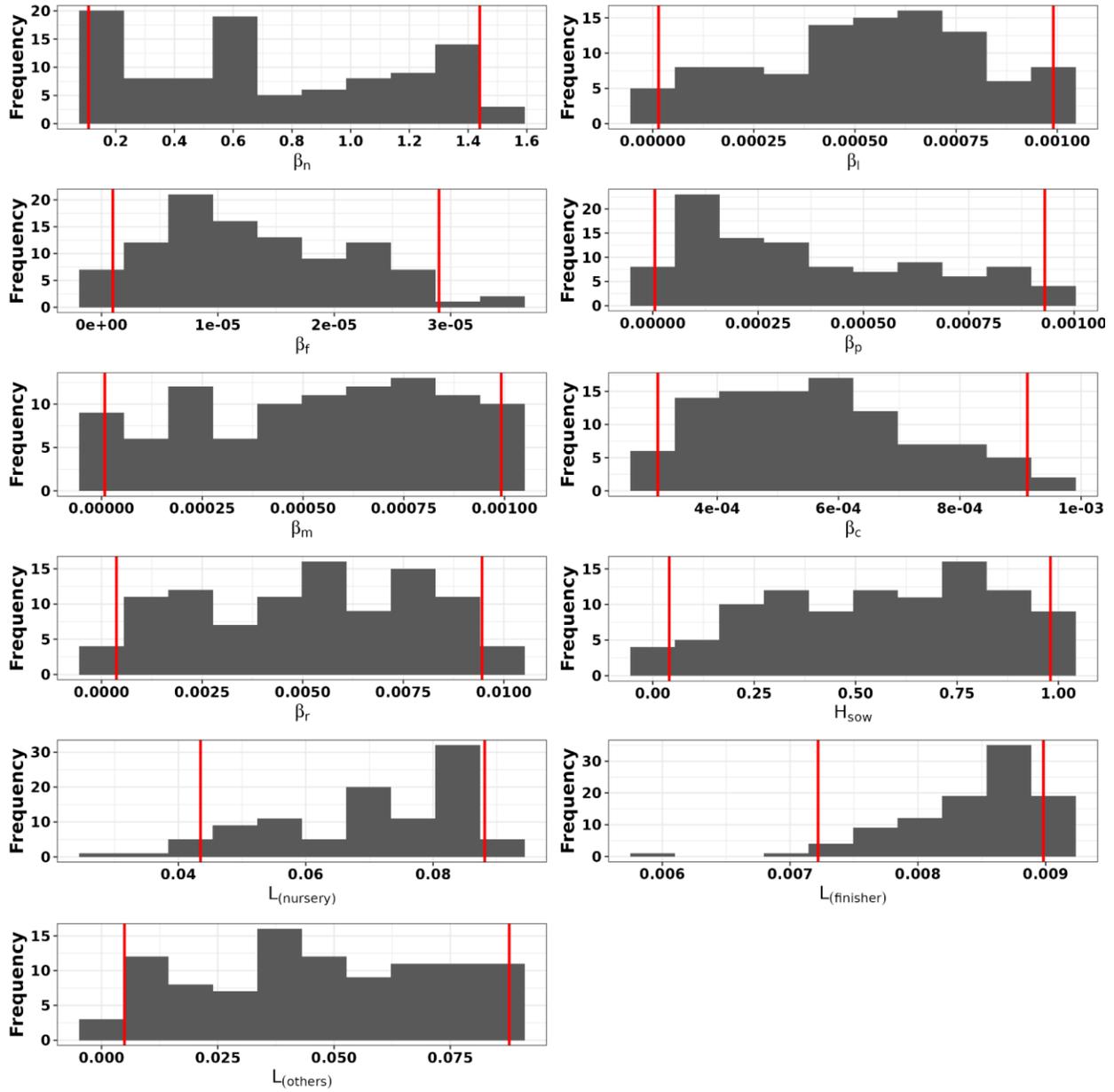

**Figure S6.** Posterior distribution of the calibrated transmission parameters derived from 100 accepted particles, red lines represent 95% credible intervals calculated through equal-tailed interval (ETI) method.

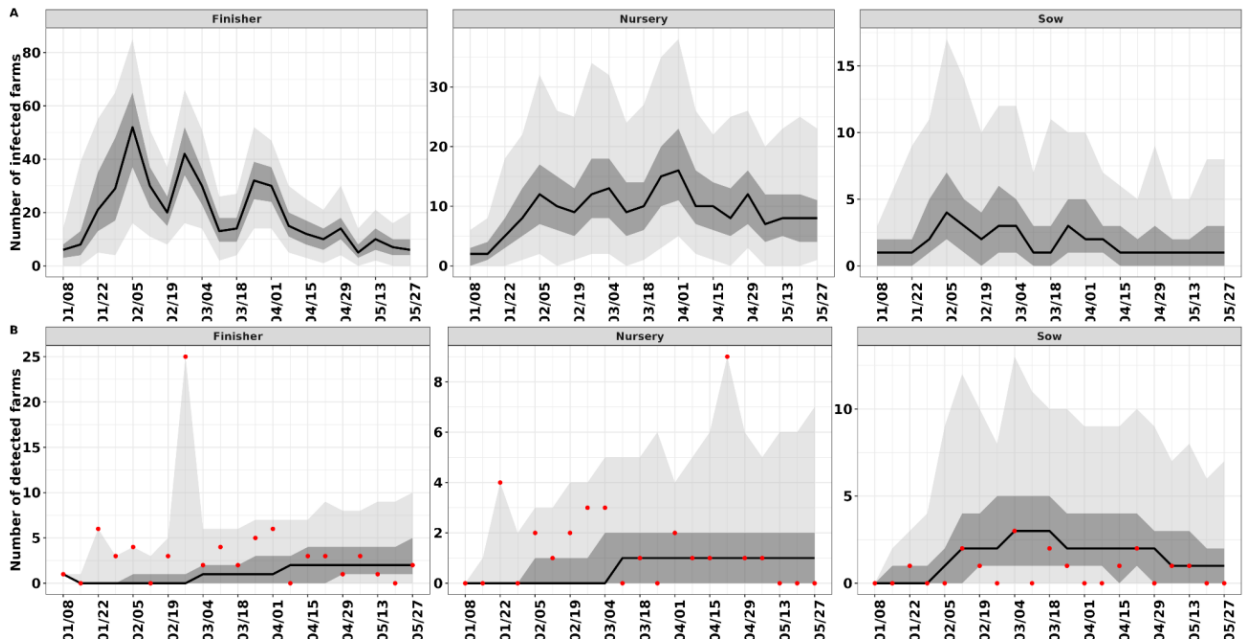

**Figure S7**. The simulated weekly number of infected farms in A) and infected detected farms (PEDV outbreaks) in B). The black line represents the median, the dark shade areas represent a 75% credible interval and the light shade areas maximum and minimum generated by the model, and the red dots the frequency of true outbreaks reported in our data. Uncertainty in the estimated model parameters is reflected by 1,000 repeated simulations.

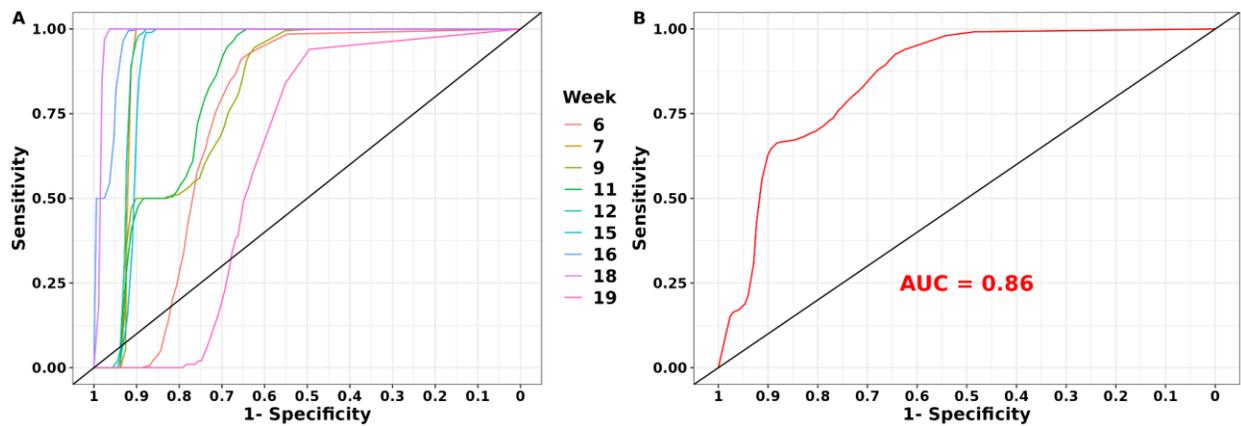

**Figure S8**. The average sensitivity and specificity for the weekly forecasts in A) and the average of all weeks in B) values calculated from 100 model calculations with each model calculation having 100 individual model iterations per predicted week to estimate the spatial location of observed outbreaks.

**Section 3**: Descriptive analysis of the between-farm pig movements and transportation vehicle movement networks, and the quantity of animal by-products in feed ingredients.

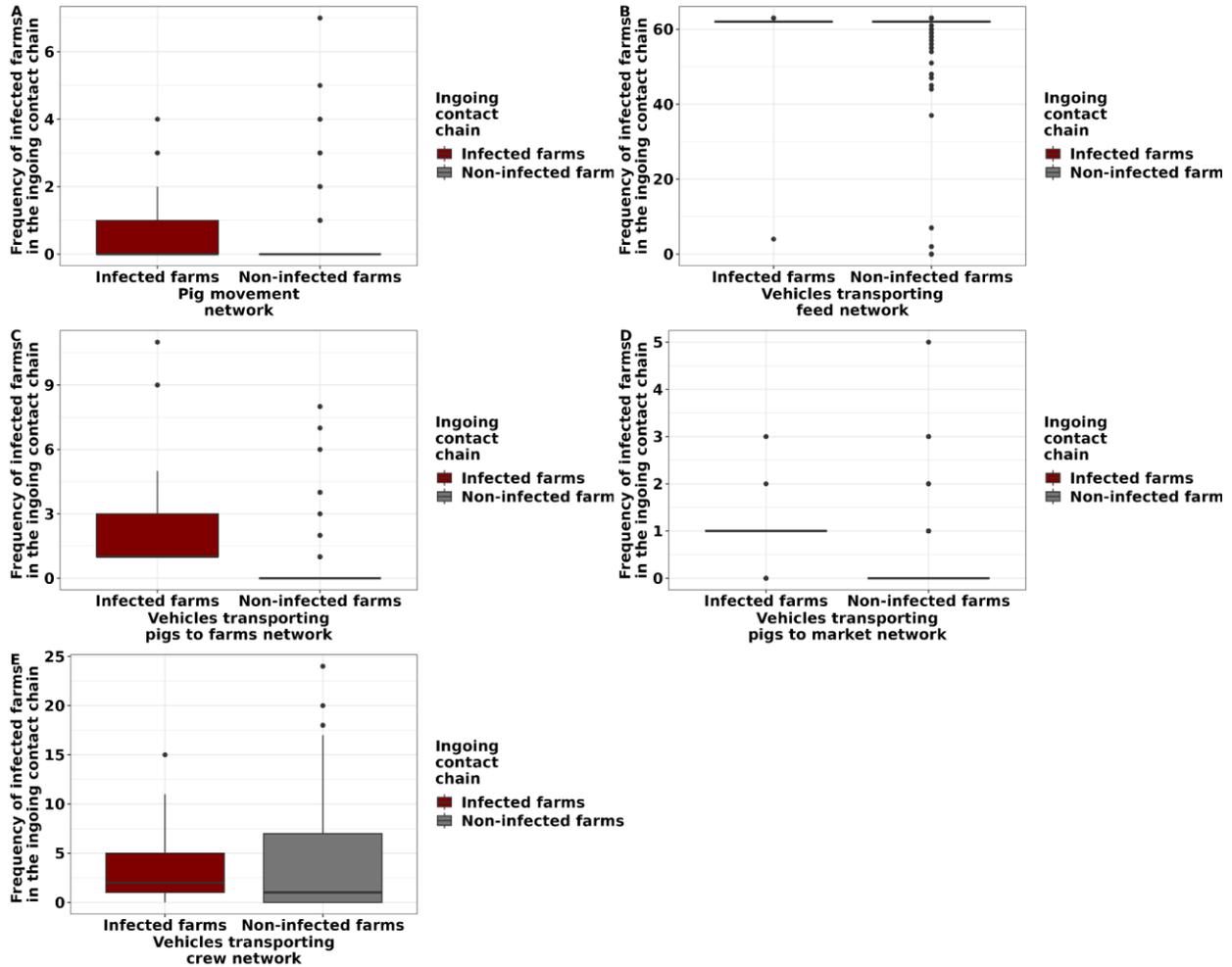

**Figure S9**. Boxplot comparing the frequency of infected farms in the ingoing contact chain of infected and non-infected farms of each transportation vehicle and pig movement network. Infected farms are more frequent in the ingoing contact chain of other infected farms for all networks (Mann Whitney test $p < 0.05$).

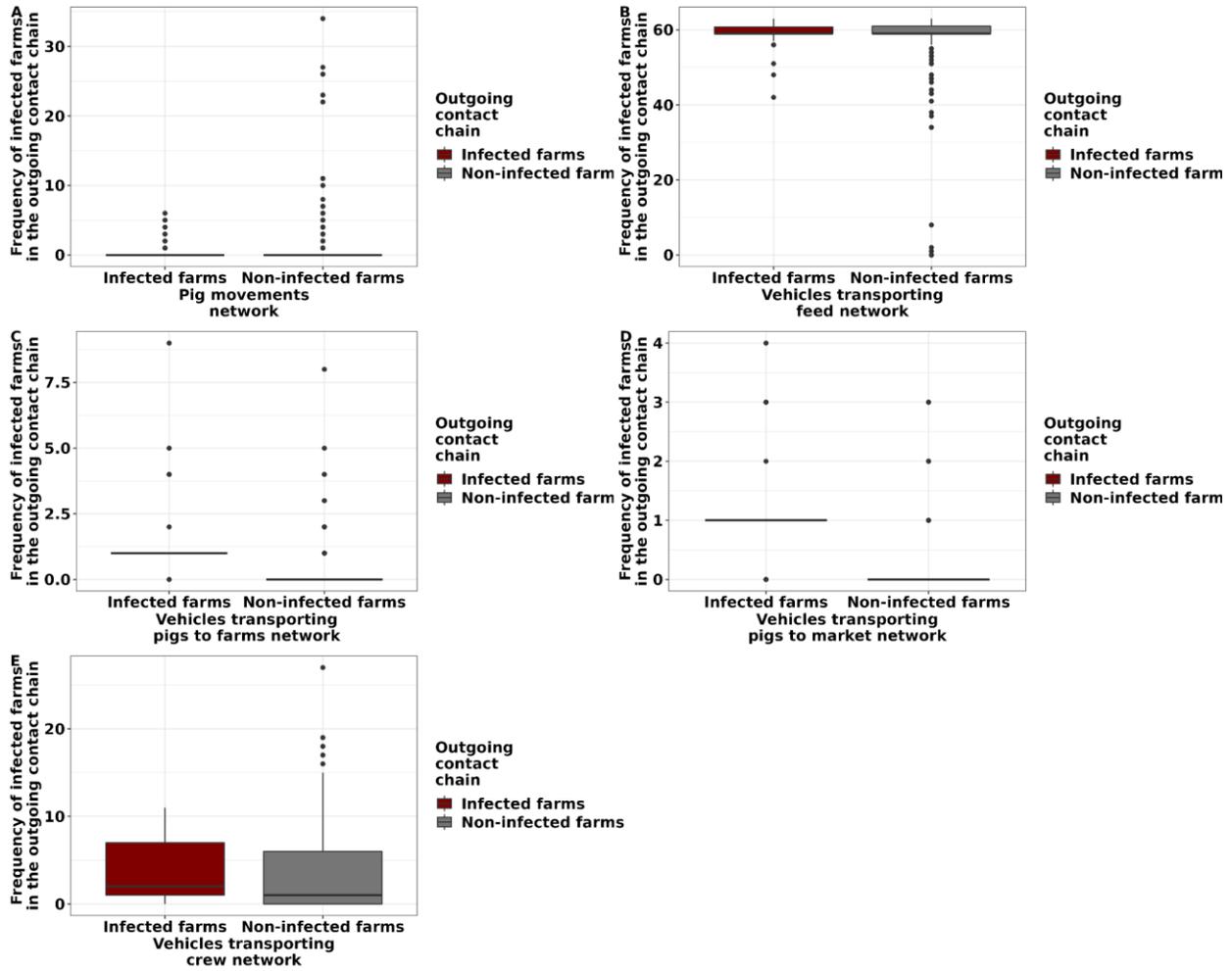

**Figure S10**. Boxplot comparing the frequency of infected farms in the outgoing contact chain of infected and non-infected farms of each transportation vehicle and pig movement network. Infected farms are more frequent in the outgoing contact chain of other infected farms for pig movements, vehicles transporting pigs to farms, vehicles transporting pigs to market, and vehicles transporting crew (Mann Whitney test $p < 0.05$).

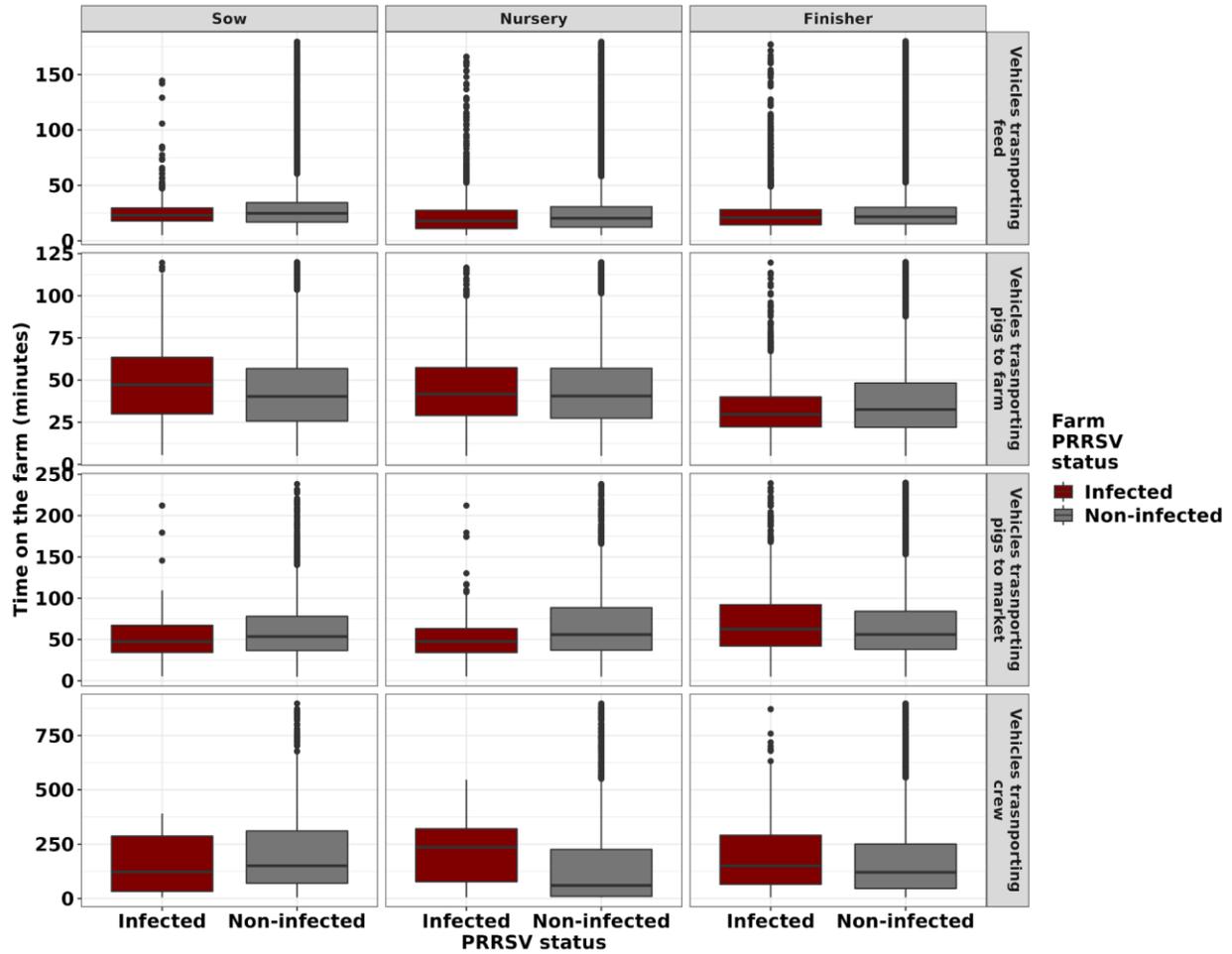

**Figure S11**. Boxplots compare the time vehicles remain on the farms of infected and non-infected farms for the different transportation vehicles (rows) and production types (columns). Vehicles transporting pigs to farms in nursery farms, vehicles transporting pigs to market in finisher farms, and vehicles transporting crew in nursery and finisher were the only vehicles that showed a higher average of time on infected farms (Mann Whitney test $p < 0.05$).

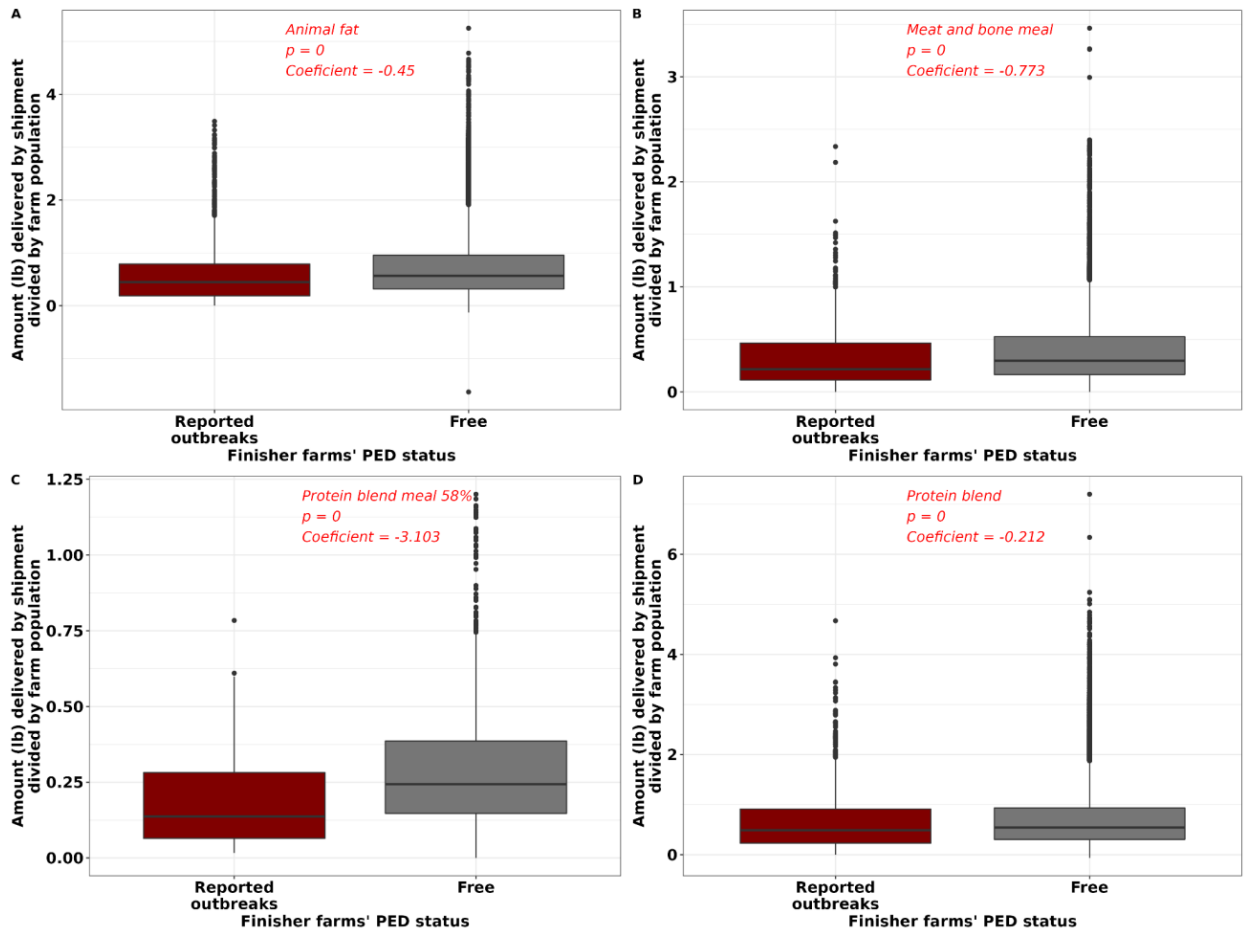

**Figure S12.** Boxplot comparing the distribution of **A**) animal fat, **B**) meat and bone meal, **C**) Protein blend meal 58% and **D**) protein blend in the feed meal received by the finisher farms with and without PEDV records in 2020 (in red the result from the logistic regression).

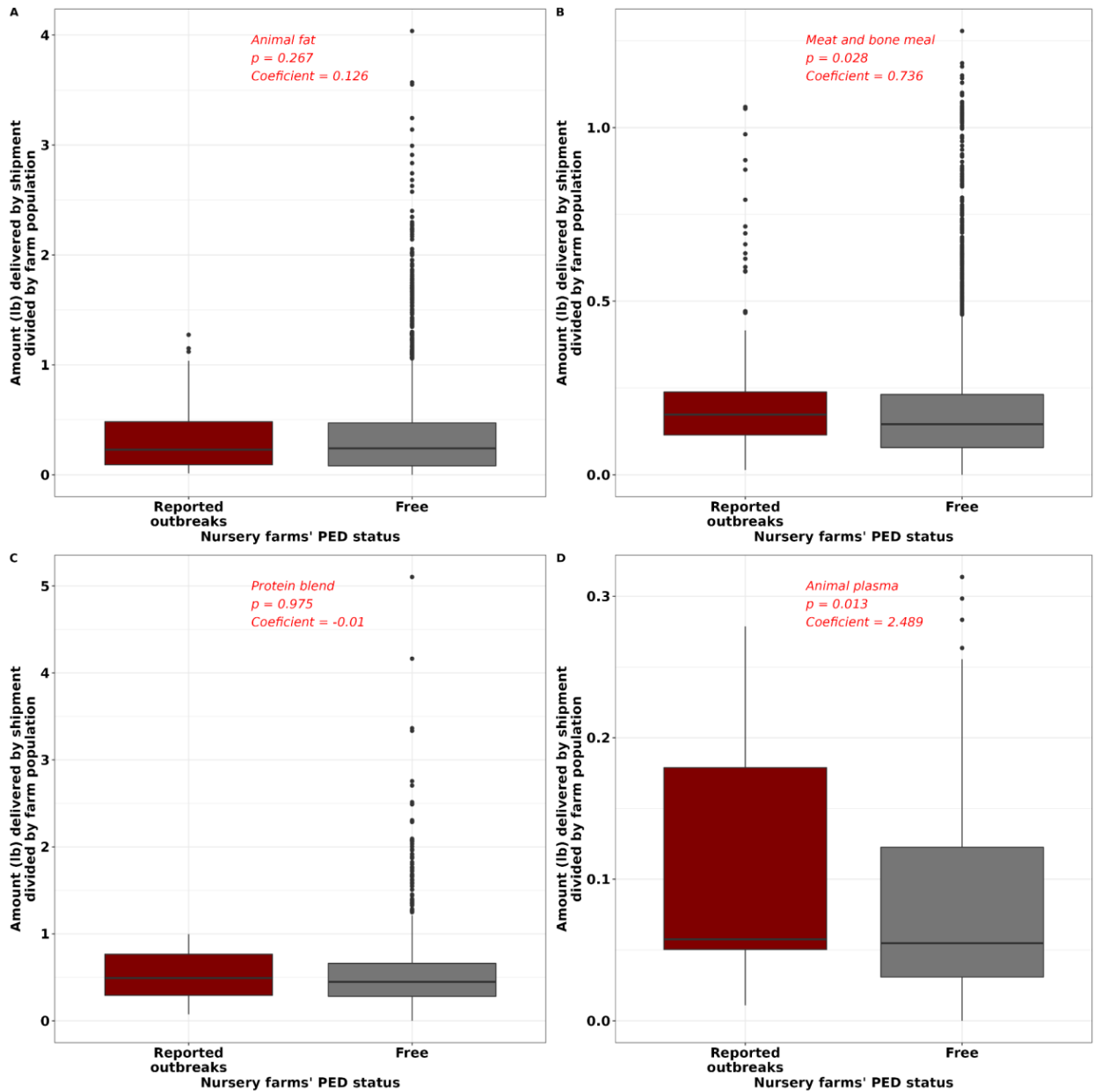

**Figure S13.** Boxplot comparing the distribution of **A**) animal fat and **B**) meat and bone meal, **C**) Protein blend and **D**) plasma in the feed meal received by the nursery farms with and without PEDV records in 2020 (in red the result from the logistic regression).

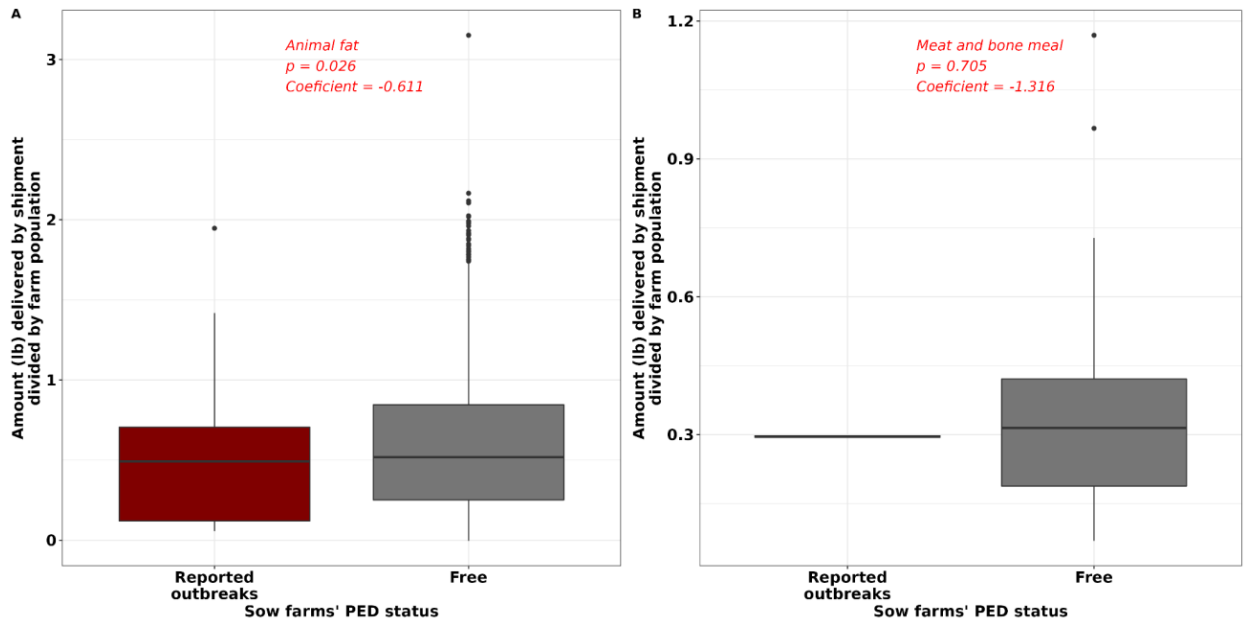

**Figure S14.** Boxplot comparing the distribution of **A**) animal fat and **B**) meat and bone meal in the feed meal received by the sow farms with and without PEDV records in 2020 (in red the result from the logistic regression).

*Section 4: contribution of the transmission routes.*

For company B, re-break was the main source of farm infections in sow farms with a contribution of 52% (95% CI 0.8%-94%) to PEDV transmission, followed by local transmission with 26.4% (95% CI 0%-93%) and pig movements with 21.6% (95% CI 0%-81%), for nursery 51% (95% CI 0%-96%) of PEDV transmission was related to pig movements and 49% (95% CI 4%-100%) to local transmission, while in finishers 70% (95% CI 12%-100%) was related to local transmission and 30% (95% CI 0%-88%) to pig movements (Figure S15). Finally, for farms of company C, re-break was the most important transmission route for sow farms contributing with 50.9% (95% CI 19%-71%) of the farm infections, followed by local transmission with 45.6% (95% CI 19%-66%) and pig movements 3.5% (95% CI 0%-37%), in nursery 100% was related to local transmission, and for finishers 89% (95% CI 24%-100%) was related to local transmission and 11% (95% CI 0%-76%) to pig movements (Figure S15).

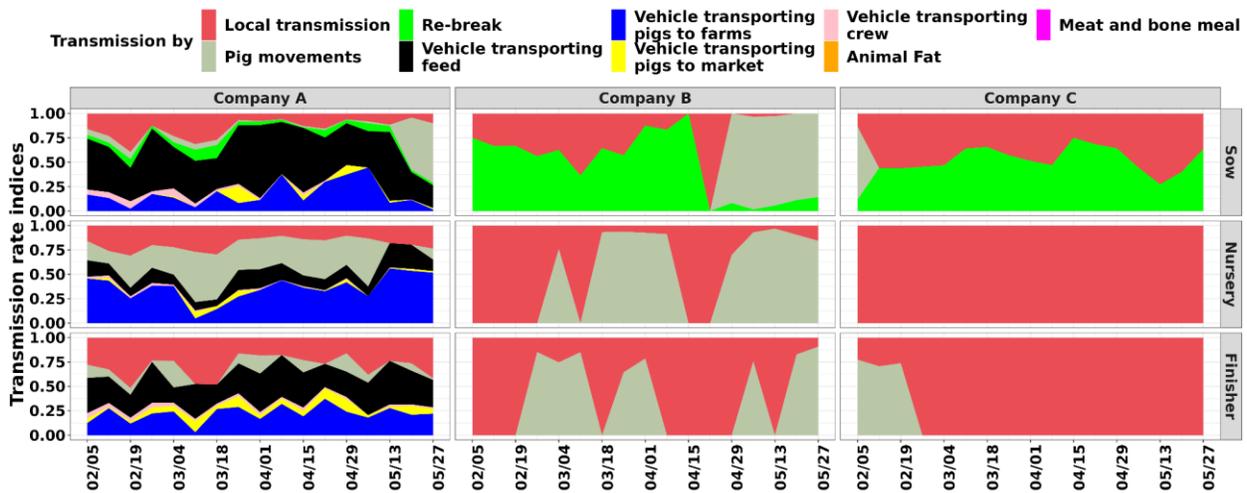

**Figure S15. Farm infection contribution for each transmission route of each company (columns) and farm types (rows).** The *y*-axis represents the proportion of transmission by each transmission route, while the x-axis shows each week in the simulation. Weekly proportions of transmission were calculated by dividing the number of simulated infected farms for each transmission route by the number of simulated infected farms by the total number of routes combined. White areas represent weeks without farm infections. Of note, animal fat and meat and bone meal contributions were at zero percent.